\documentstyle[12pt]{article}

\catcode`\@=11
\long\def\@makefntext#1{
\protect\noindent \hbox to 3.2pt {\hskip-.9pt  
$^{{\ninerm\@thefnmark}}$\hfil}#1\hfill}                

\def\@makefnmark{\hbox to 0pt{$^{\@thefnmark}$\hss}}  
        
\def\ps@myheadings{\let\@mkboth\@gobbletwo
\def\@oddhead{\hbox{}
\rightmark\hfil\ninerm\thepage}   
\def\@oddfoot{}\def\@evenhead{\ninerm\thepage\hfil
\leftmark\hbox{}}\def\@evenfoot{}
\def\sectionmark##1{}\def\subsectionmark##1{}}

\renewcommand{\thefootnote}{\fnsymbol{footnote}}

\newcounter{sectionc}\newcounter{subsectionc}\newcounter{subsubsectionc}
\renewcommand{\section}[1] {\vspace*{0.6cm}\addtocounter{sectionc}{1} 
\setcounter{subsectionc}{0}\setcounter{subsubsectionc}{0}\noindent 
        {\normalsize\bf\thesectionc. #1}\par\vspace*{0.4cm}}
\renewcommand{\subsection}[1] {\vspace*{0.6cm}\addtocounter{subsectionc}{1} 
        \setcounter{subsubsectionc}{0}\noindent 
        {\normalsize\it\thesectionc.\thesubsectionc. #1}\par\vspace*{0.4cm}}
\renewcommand{\subsubsection}[1]
{\vspace*{0.6cm}\addtocounter{subsubsectionc}{1}
        \noindent
{\normalsize\rm\thesectionc.\thesubsectionc.\thesubsubsectionc. 
        #1}\par\vspace*{0.4cm}}

\newcounter{appendixc}
\newcounter{subappendixc}[appendixc]
\newcounter{subsubappendixc}[subappendixc]

\renewcommand{\appendix}[1] {\vspace*{0.6cm}
        \refstepcounter{appendixc}
        \setcounter{figure}{0}
        \setcounter{table}{0}
        \setcounter{equation}{0}
        \renewcommand{\thefigure}{\Alph{appendixc}.\arabic{figure}}
        \renewcommand{\thetable}{\Alph{appendixc}.\arabic{table}}
        \renewcommand{\theappendixc}{\Alph{appendixc}}
        \renewcommand{\theequation}{\Alph{appendixc}.\arabic{equation}}
        \noindent{\bf Appendix \theappendixc #1}\par\vspace*{0.4cm}}

\renewenvironment{thebibliography}[1]
        {\begin{list}{\arabic{enumi}.}
        {\usecounter{enumi}\setlength{\parsep}{0pt}
\setlength{\leftmargin 1.25cm}{\rightmargin 0pt}
         \setlength{\itemsep}{0pt} \settowidth
        {\labelwidth}{#1.}\sloppy}}{\end{list}}

\newcounter{itemlistc}
\newcounter{romanlistc}
\newcounter{alphlistc}
\newcounter{arabiclistc}

\newcommand{\fcaption}[1]{
        \refstepcounter{figure}
        \setbox\@tempboxa = \hbox{\footnotesize Fig.~\thefigure. #1}
        \ifdim \wd\@tempboxa > 6in
           {\begin{center}
        \parbox{6in}{\footnotesize\baselineskip=12pt Fig.~\thefigure. #1}
            \end{center}}
        \else
             {\begin{center}
             {\footnotesize Fig.~\thefigure. #1}
              \end{center}}
        \fi}

\newcommand{\tcaption}[1]{
        \refstepcounter{table}
        \setbox\@tempboxa = \hbox{\footnotesize Table~\thetable. #1}
        \ifdim \wd\@tempboxa > 6in
           {\begin{center}
        \parbox{6in}{\footnotesize\baselineskip=12pt Table~\thetable. #1}
            \end{center}}
        \else
             {\begin{center}
             {\footnotesize Table~\thetable. #1}
              \end{center}}
        \fi}

\def\@citex[#1]#2{\if@filesw\immediate\write\@auxout
        {\string\citation{#2}}\fi
\def\@citea{}\@cite{\@for\@citeb:=#2\do
        {\@citea\def\@citea{,}\@ifundefined
        {b@\@citeb}{{\bf ?}\@warning
        {Citation `\@citeb' on page \thepage \space undefined}}
        {\csname b@\@citeb\endcsname}}}{#1}}

\newif\if@cghi
\def\cite{\@cghitrue\@ifnextchar [{\@tempswatrue
        \@citex}{\@tempswafalse\@citex[]}}
\def\citelow{\@cghifalse\@ifnextchar [{\@tempswatrue
        \@citex}{\@tempswafalse\@citex[]}}
\def\@cite#1#2{{$\null^{#1}$\if@tempswa\typeout
        {IJCGA warning: optional citation argument 
        ignored: `#2'} \fi}}

 1
 1
 1

\font\ninerm=cmr9

\begin{document}
\input psfig

\centerline{\normalsize\bf APPLICATION OF CURRENT ALGEBRA OR CHIRAL SYMMETRY}
\baselineskip=22pt
\centerline{\normalsize\bf TO TAU HADRONIC DECAYS\footnote{
 Talk given at  International Symposium on Heavy Flavour and Electroweak Theory (Beijing, 1995)} }
\baselineskip=16pt
\centerline{\normalsize\bf }

\centerline{\footnotesize L. BELDJOUDI}
\baselineskip=13pt
\centerline{\footnotesize\it Centre de Physique Th\'eorique,
 Ecole Polytechnique}
\baselineskip=12pt
\centerline{\footnotesize\it 91128 Palaiseau, France}
\centerline{\footnotesize E-mail: beljoudi@orphee.polytechnique.fr}
\vspace*{0.3cm}
\centerline{\footnotesize and}
\vspace*{0.3cm}
\centerline{\footnotesize T. N. TRUONG}
\baselineskip=13pt
\centerline{\footnotesize\it Centre de Physique Th\'eorique,
Ecole Polytechnique}
\centerline{\footnotesize E-mail: truong@orphee.polytechnique.fr}
\vspace*{0.9cm}

\abstract{$\tau\to\pi\pi\nu$, $\tau\to\pi K\nu$, $\tau\to K\eta\nu$,
 $\tau\to 3\pi\nu$ and $\tau\to \pi\pi K\nu$ have been investigated
 using chiral symmetry with dispersion relation in agreement with the
 unitarity condition.}
 
\normalsize\baselineskip=15pt
\setcounter{footnote}{0}
\renewcommand{\thefootnote}{\alph{footnote}}
\section{Introduction}
        Current Algebra was invented a long time ago to study
 phenomena involving emission of soft pions (and kaons) and also the
 chiral symmetry breaking effects.  It was soon discovered that the
 current algebra low energy theorems (LET) can be obtained by
the effective lagrangian method at the tree graph order.  Using
it, we can calculate matrix elements of physical processes by the Feynman
perturbation method in a straightforward way. Because of the chiral
properties of the Nambu-Goldstone bosons, the effective lagrangian involves the
derivatives of the pion fields which make the theory non-renormalisable.
        Chiral Perturbation Theory ($\chi$PT) is a well defined perturbation
 procedure which can take into account in a systematic way of higher orders
 (higher
loops) and of the chiral symmetry breaking effects. At any order, these relations form low energy theorems
of QCD. Despite
 the beauty of the method, it has limitations in phenomenological
 applications. We distinguish  three
main problems with this theory:

        1) Because of the non-renormalisability of the theory, the number
of parameters increases as we calculate higher loops.

        2) $\chi$PT is effectively a power series expansion in
 terms of pion momenta,
 hence it
cannot take into account of the resonance effects (Breit Wigner
 form cannot be
expanded in a convergent power series  of momenta). It can only explain the 
low energy tail of the resonance.

        3) Because $\chi$PT is a perturbation theory, it only satisfies the
unitarity relation order by order. This approximation is not good enough
for strong interaction or for resonance physics.

        Because of these problems, at first sight, $\chi$PT cannot be used to
 study Tau physics where most of the  hadronic modes are dominated by resonances and
 where the pions emitted are not soft. The
remedy for these problems were given a long time ago even before $\chi$PT
 became
popular. 
        One method is to use the Current Algebra soft pion theorems and use
dispersion relation to take into account of the  unitarity. The second
 method is to extend
the validity of the $\chi$PT by resumming the perturbation series in order to
take into account of the unitarity relation. The third method is to assume
the vector meson dominance in the electroweak  processes
 and
requiring at low energy or at zero momentum transfer the low enegy 
theorem is recovered. In fact, the three methods are equivalent as they all
satisfy the LET and the unitarity relation to a  good approximation.
  Using the basic Equal Time Commutator Relations (ETCR) as postulated by
 Gell-Mann,
\begin{eqnarray}
\left[Q_{V}^{a},V^{b}_{\mu}(x)\right]&=&if^{abc}V^{c}_{\mu}(x)\\
\left[Q_{A}^{a},A^{b}_{\mu}(x)\right]&=&if^{abc}V^{c}_{\mu}(x)\\
\left[Q_{V}^{a},A^{b}_{\mu}(x)\right]&=&if^{abc}A^{c}_{\mu}(x)
\end{eqnarray}                   
$V^{a}_{\mu}$ and $A^{a}_{\mu}$ are the vector and axial currents
 generated by the
approximate symmetry $SU(3)_L\times SU(3)_R$ of QCD, $Q_{V}^{a}$ and
$Q_{A}^{a}$ are the corresponding charges.
 
 Together with the basic Current Algebra  formula for a soft pion
emission :
 \begin{equation}
\lim_{k_{\mu}\rightarrow 0} \langle\pi^{a}(k) B\vert
O(0)\vert A\rangle =-{i\over f_\pi} \langle
B\vert\left[Q_A^{a}, O(0)\right]\vert A\rangle
\end{equation}

Eq(4) sets the scale for the relevant matrix element. In the dispersion
 relation
 language it can be used as a substraction constant for a substracted dispersion relations.

More explicitly, let us denote the Current Algebra result as
 $A_{LET}(0)$, thus the substracted dispersion relation can be written as
follows:
\begin{equation}
A(s)= A_{LET}(0)+{s\over\pi}\int\limits_{s_0}^\infty
dz{{\rm{Im}}\ A(z)\over
z(z-s-i\epsilon)}
\end{equation} 

Because of the substracted dispersion relation, the high energy  contribution in the dispersion integral, which is difficult to calculate reliably,
 is suppressed. This result is in contrast with the physics done in the 60's
 (e.g. Bootstrap) which involves unsubstracted dispersion relation with
 uncontrollable  approximation (because the high energy contribution is not
suppressed).
       
In the physical world, where the pion are not soft, corrections must be
made.  First the chiral symmetry breaking effect can be
taken into account approximately  by hand. The second
correction to be made is to impose unitarity relation which is crucial for
hadronic processes. This can be done either by strong interaction dynamics
models e.g. Vector Meson Dominance or some unitarisation schemes.
       
Some of these relations were calculated in 1978\cite{prehist}.  It is desirable
to update these calculations with a more sophisticated technique in order
to compare with more recent experimental data and also to the $\chi$PT
technique. As an illustration we present applications to $\tau\to\pi\pi\nu$,
$\tau\to\pi K\nu$, $\tau\to K\eta\nu$ $\tau\to\pi\rho\nu$, 
$\tau\to\pi K^*\nu$ and 
$\tau\to K\rho\nu$ decays. 

\section{$\tau\to\pi\pi\nu$ decay}

Using the CVC hypothesis and Lorentz invariance, one can write
straightforwardly 
the following matrix elements:
\begin{eqnarray}
&\langle\pi^{a}(p_1) \pi^{b}(p_2)\vert&V_{\mu}^{c} (0)\vert0
\rangle =i\epsilon ^{abc}f(s)(p_2-p_1)_\mu  \\   
&\langle\pi^{a}(p_1) \pi^{b}(p_2)\vert&\hat {m}(\bar {u}u+\bar {d}d)\vert0
\rangle =\delta^{ab}\Gamma (s)  
\end{eqnarray}
At zero momentum transfer  we have the following normalization: $f(0)=1$ and
$\Gamma (0)=m_{\pi}^{2}$. 
  Assuming the elastic
unitarity condition, we deduce the following relations:
\begin{eqnarray}
 Imf(s)=f(s)\exp {-i\delta_1^{1} }\sin {\delta_1^{1}}\nonumber\\ 
Im\Gamma(s)=\Gamma(s)\exp {-i\delta_0^{0} }\sin {\delta_0^{0}}
\end{eqnarray}

Hence $f(s)$ must have the phase of the P wave $\pi\pi$ phase shift, and
 $\Gamma(s)$ the
S wave phase $\pi\pi$ shift.

The general solutions to this equation are well known, they are of the 
Muskhelishvili Omn\`es type\cite{omnes}:
\begin{eqnarray}
f(s)&=&P_{f}(s)\Omega_1 (s)\\ 
\Gamma(s)&=&\Gamma(0)P_{\Gamma}(s)\Omega_0 (s)
\end{eqnarray}
where 
\begin{eqnarray}
\Omega_1 (s)&=&\exp ({{s\over \pi}\int\limits_{4m_\pi^2} ^{+\infty} {
\delta_1^{1}dz\over z(z-s-i\epsilon}})\\
\Omega_0 (s)&=&\exp ({{s\over \pi}\int\limits_{4m_\pi^2} ^{+\infty} {
\delta_0^{0}dz\over z(z-s-i\epsilon}})
\end{eqnarray}

$P_{f}$ and $P_{\Gamma}$ are polynomials normalized to unity at $s=0$
 which  determine the high energy
behavior of the form factors. They could also represent the low energy
contribution of the higher
mass intermediate states to the form
factors. In the following we assume the dominance of
the elastic unitarity relation and hence we set
$ P_{f}(s)=P_{\Gamma}(s)=1$.

At one loop chiral perturbation theory we have:
\begin{eqnarray}
f(s)\hskip -2mm&=&\hskip -4mm1+{r_v^2\over 6}s-(96\pi f_{\pi}^2)^{-1}\hskip -2mm\left[
(s-4m_{\pi}^2)(h(s)-h(0))+4m_{\pi}^2h'(0)s\right]\\
{\Gamma(s)\over\Gamma(0)}\hskip -2mm&=&\hskip -4mm 1+{r_s^2\over 6}s-(16\pi f_{\pi}^2)^{-1}\hskip -2mm\left[
(s-{m_\pi^2\over 2})(h(s)-h(0))+{m_\pi^2\over 2}sh'(0)\right]
\end{eqnarray}

If we make a bubble summation consistently with the elastic unitarity
condition, we obtain:
\begin{eqnarray}
f(s)={1\over 1-{r_v^2\over 6}s+(96\pi f_{\pi}^2)^{-1}\left[
(s-4m_{\pi}^2)(h(s)-h(0))+4m_{\pi}^2h'(0)s\right] }\nonumber\\
\Gamma(s)={\Gamma(0)\over 1-{r_s^2\over 6}s+(16\pi f_{\pi}^2)^{-1}\left[
(s-{m_\pi^2\over 2})(h(s)-h(0))+{m_\pi^2\over 2}sh'(0)\right]}
\end{eqnarray}
where 
\begin{equation}
h(s)={1\over\pi}\rho(s)ln({1+\rho(s)\over -1+\rho(s)})
\end{equation}
$\rho(s)$ is the phase space factor.

Eq(13) can also be obtained by the diagonal Pad\'e approximant method by using
Eq(11) and (12).
This approximation is identical to the Vector Meson Dominance approach of
Gell-Mann-Sharp and Wagner where the self energy correction is taken into
account to make the vector meson instable and with $\rho\pi\pi$ coupling given
by KSRF relation\cite{ksrf}.

An alternative method to calculate  the scalar and
vector form factors is to use the S and P wave $\pi\pi$ phase shifts and the
 Omn\`es representation (see Eq.(9) and Eq.(10)). For more details
 see Ref.\cite{belz}. The P wave $\pi\pi$ phase
shift is given in Fig.(1).

\begin{figure}
{\psfig{figure=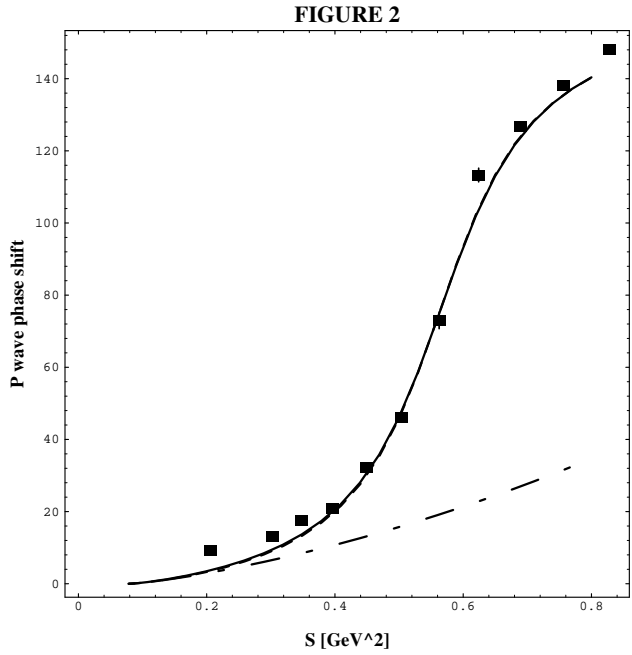,height=8.5truecm,angle=0}}
\caption{\em The solid line $I=1$, $l=1$ pion scattering phase shift
calculated from
unitarized ChPT Dashed line corresponds to the similar phase shift where the
left hand cut is neglected.}
\end{figure}

 Because there are no experimental
information on the
scalar form factor, we only compare the vector form factor with the
experimental data. It is
seen that the agreement between theory and experimental data is
satisfactory although the peak
values of the experimental form factor squared at the $\rho$ mass is about
40 as compared with
the theoretical calculation value 32 or an error of the order
of 20\% as it can be seen in Fig.(2).

\begin{figure}
{\psfig{figure=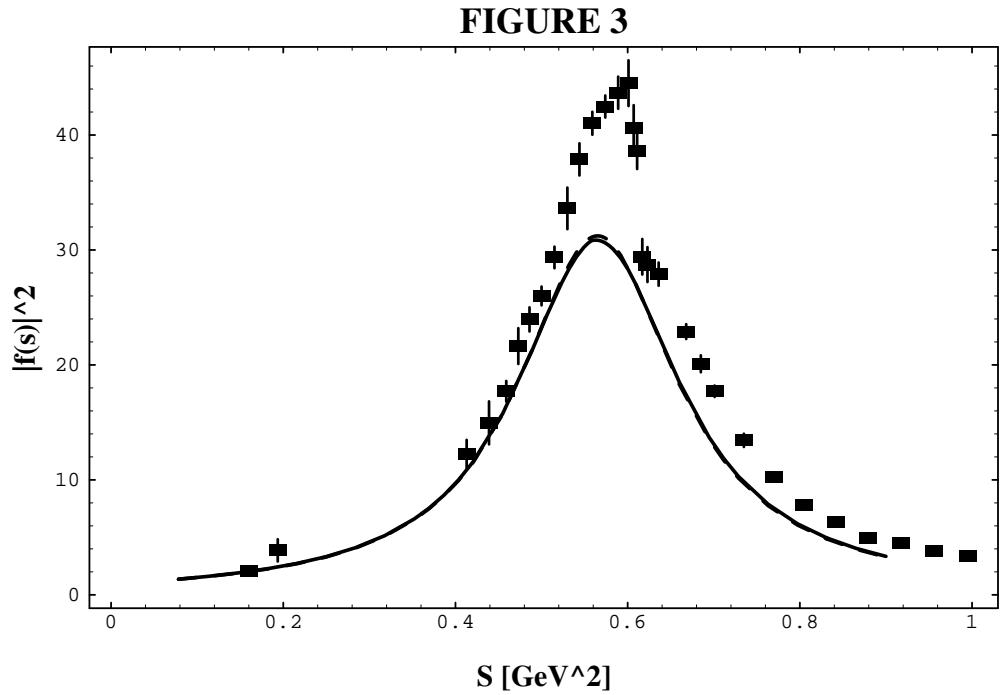,height=9truecm,angle=0}}
\caption{\em The solid line is the  vector pion form factor squared 
calculated from the
Omn\'es representation. The dashed line represents the same quantity
calculated by the Pad\'e Method}
\end{figure} 

 This
discrepancy is 
probably due to the inelastic effect of the $\omega \pi$ channel as was
previously pointed
out\cite{costa}.

We can also calculate the vector and scalar pion rms radii using the
following formula:
 $$\langle r^2_{V}\rangle={6\over \pi} \int\limits_{4m_\pi^2} ^{+\infty} {
\delta_1^{1}dz\over z^2} $$

$$\langle r^2_{S}\rangle={6\over \pi} \int\limits_{4m_\pi^2} ^{+\infty} {
\delta_0^{0}dz\over z^2 }$$
Numerical integration gives $\langle r^2_{V}\rangle=
0.40 fm^2$, and  $\langle r^2_{S}\rangle = 0.47 fm^2$ compared to the
experimental value $\langle r^2_{V}\rangle = 0.439\pm 0.03 fm^2$, 
and the $\langle r^2_{S}\rangle=0.5 fm^2 $ which is obtained from the
experimental $\pi K$ 
scalar radius and from SU(3) symmetry. The agreement between experimental
data and theoretical
 calculations is  satisfactory.

\section{$\tau\to\pi K\nu$ decay}

The most general $\tau \to \pi K \nu$ decay amplitudes are given in terms of
two form factors: 
\begin{equation}
\langle\pi ^{0} K^{-}\vert V_{\mu} ^{4-i5}(0)\vert0
\rangle=f_1(s)(p_2-p_1)_\mu+f_2(s)(p_1+p_2)_\mu 
\end{equation}
where $ p_1$, and $ p_2$ are, respectively,  the  pion and kaon momenta,
and $ s=(p_1+p_2)^2$ is
the time-like  momentum transfer  and  $V_{\mu}^{4-i5}$ is the vector
current operator with the
superscript indices referring to the SU(3) octet currents.  
 $f_1(s)$ is the P wave $\pi K$ form factor,  $f_2(s)$ is a linear
combination of S and P states as can be
seen by taking the divergence of Eq (15):
\begin{equation}
 g(s)= -i \langle \pi ^{0} K^{-}
\vert \partial ^{\mu} V_{\mu}^{4-i5}(0) \vert 0 \rangle =
(m_K^2-m_\pi^2)f_1(s)+sf_2(s)
\end{equation}
 g(s) is therefore a pure scalar
which describes the S wave $\pi K$ form factor.  g(s) measures the $SU(3)$
violating effect because, in the exact SU(3) limit, the vector current is
conserved. We expect therefore in the  $\tau
\to \pi K \nu$ decay, the P wave form factor $f_1(s)$ dominates.

Because of the octet current hypothesis the two channels $\pi^0 K^-$ and $\pi^-
\bar{ K^0}$ matrix elements are related by the  Clebsh Gordon coefficient
\begin{equation}
\langle\pi^- \bar K^0\vert V_{\mu} ^{4-i5}(0)\vert0\rangle=\sqrt{2}
\langle\pi^0  K^-\vert V_{\mu} ^{4-i5}(0)\vert0\rangle
\end{equation}

 Using the Ademollo-Gatto theorem $f_1(0)={1/ \sqrt 2}$ and hence\hfill\break
 $g(0) ={ (
m_K^2-m_\pi^2)/ \sqrt 2}$ for $\pi^0 K^-$ system.

Using the standard  current algebra technique and the $SU(2)_L \times SU(2)_R $
commutation relation by taking the pion momentum $p_1$ soft we have the
well known Callan-Treiman relation\cite{treiman}:
\begin{equation}
f_1(m_K^2) +f_2(m_K^2) ={ f_K \over f_\pi \sqrt 2}
\end{equation}
where $f_K$ and  $f_\pi$ are, respectively, the K and $\pi$ decay constants 
 \\${f_K/f_\pi}=1.22$.

By evaluating Eq(16) at $t=m_K^2$ and noting that $f_2(m_K^2)$ is proportional
to ${ m_{\pi}^2/ m_{K}^2 }$ we
have: $g(m_K^2) \approx   g(0){ f_K/f_\pi}$ 

The elastic unitarity condition, which should be valid in the physical
region of the
$\tau\to\pi K\nu$ decay gives:
\begin{eqnarray}
 Imf_1(s)&=f_1(s)\exp {-i\delta_p^{1/2} }\sin {\delta_p^{1/2}}\\
 Img(s)&=g(s)\exp {-i\delta_s^{1/2} }\sin {\delta_s^{1/2}}
\end{eqnarray}
where $\delta_s^{1/2}$ and $\delta_p^{1/2}$ are respectively the phase of S and
P wave I=1/2 $\pi K$ scattering amplitude. 
The general solutions are given by  the 
Muskhelishvili Omn\`es representation\cite{omnes}:
\begin{eqnarray}
f(s)=f(0)\exp ({{s\over \pi}\int\limits_{4m_\pi^2} ^{+\infty} {
\delta_p^{1/2}dz\over z(z-s-i\epsilon}})\\
g(s)=g(0)\exp ({{s\over \pi}\int\limits_{4m_\pi^2} ^{+\infty} {
\delta_s^{1/2}dz\over z(z-s-i\epsilon}})
\end{eqnarray}

At one loop chiral perturbation theory we have:
\begin{eqnarray}
f(s)=f(0)+f^{loop}(s)\\
g(s)=g(0)+g^{loop}(s)
\end{eqnarray}
$f^{loop}(s)$ and $g^{loop}(s)$ are given elsewhere\cite{belp}.
If we make the usual bubble summation, consistent with the elastic unitarity
condition, we obtain:
\begin{eqnarray}
f(s)={f(0)\over{1-f^{loop}(s)/f(0)}} \nonumber\\
g(s)={g(0)\over{1-g^{loop}(s)/g(0)}}
\end{eqnarray}

From the expression for $f_1(s)$, the phase of the form factor which is
identical to the 
P wave phase shifts of $\pi K$ scattering amplitude, can be calculated
using the experimental value of $ \langle r_v^2\rangle\ 
=0.34 \pm 0.03 fm^2$. Using this value we have $m_{K^*}=810\pm 30$ MeV which
agrees with 
the experimental data $m_{K^*}=892$MeV. Its width satisfies the following
modified
KSRF relation\cite{ksrf}:
\begin{equation}
\Gamma _{K^*}={ 
\lambda^{3/2}(m_{K^*}^2,m_\pi ^2,m_K^2) \over 128 \pi m_{K^*}^3
f\pi^2}
\end{equation}
Using the experimental value $m_{K^*}=892$MeV the numerical result of the 
right hand side of Eq(26) is $55$ MeV, 
compared to the experimental value of $49.8 \pm 0.8$ MeV. In Fig.(3), we give
a graphic representation of $f_1(s)$ squared.

\begin{figure}
{\psfig{figure=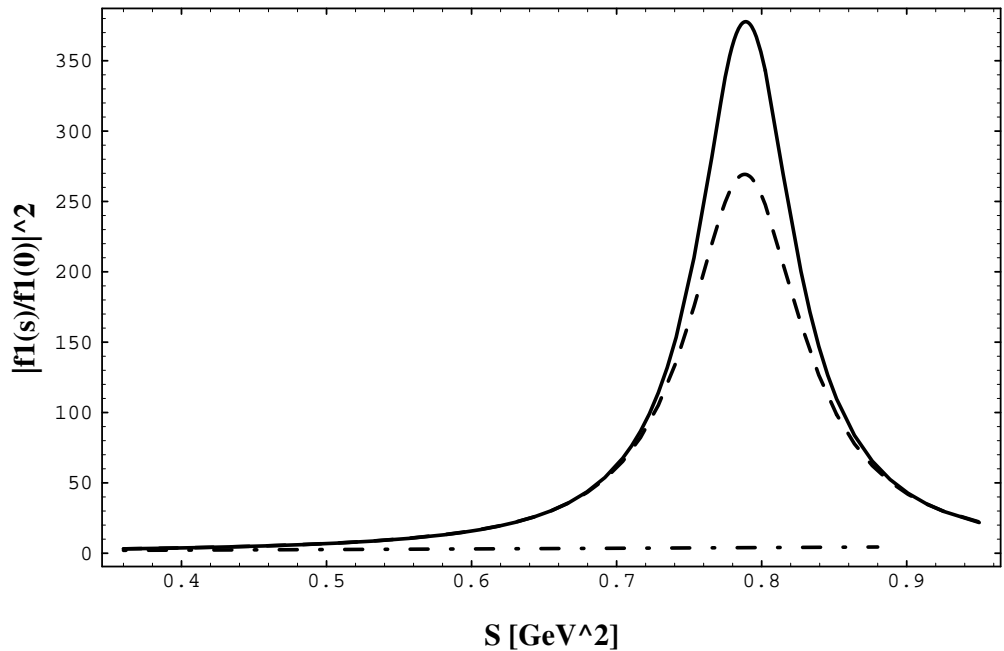,height=8.5truecm,angle=0}}
\caption{\em The calculated  P wave  $\pi K$ form factors 
(solid/dashed/dot-dashed curves) corresponding
 respectively to, the Omn\'es representation using the $\pi K$ phase shift
calculated from the 
unitarized 
one loop CPTh, the [1,1] Pad\'e approximant,
CPTh prediction }
\end{figure}

The branching ratio $B.R = {\Gamma( \tau \to \pi K \nu ) \over \Gamma(\tau
\to all)}$ is $ 1.0 \% $ and is in agreement with the experimental result
of $ B.R_{exp.}= (1.4 \pm 0.2) \% $

Because the S wave $\pi K$ scattering length does not vanish, g(s) has a
square root threshold
singularity at the threshold (the derivative of g(s) is discontinuous at
this point) as it
can be seen in Fig.(4).

\begin{figure}
{\psfig{figure=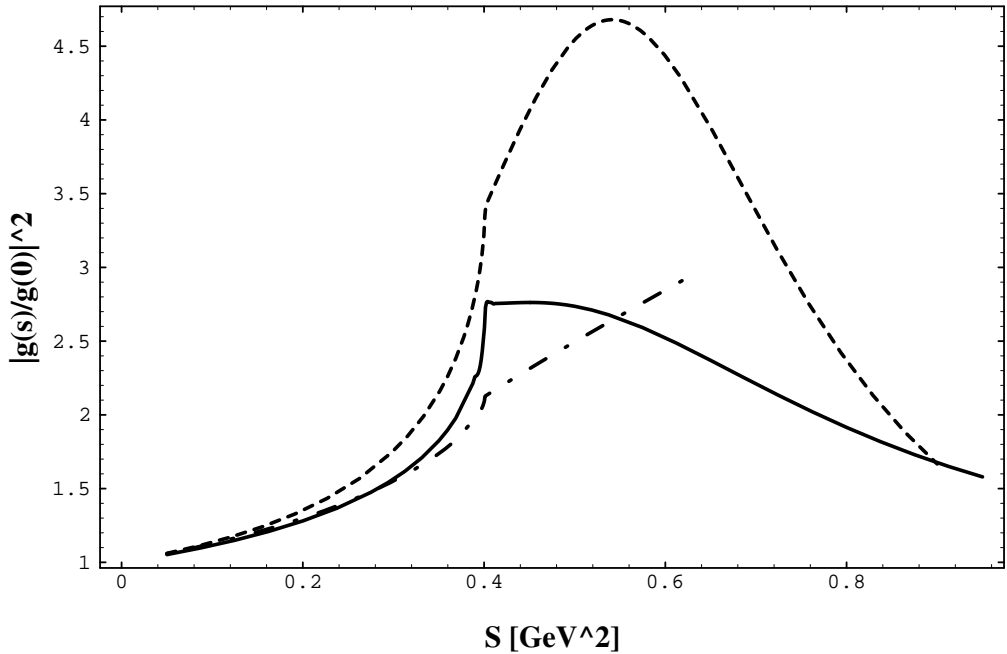,height=8.5truecm,angle=0}}
\caption{\em The calculated  S wave $\pi K$ form factors 
(solid/dashed/dot-dashed curves) corresponding
 respectively to, the Omn\'es representation using the $\pi K$ phase shift
calculated from
 unitarized one 
loop CPTh, the [1,1] Pad\'e approximant , CPTh prediction. }
\end{figure}

The scalar form factor contributes very little to the $\pi K$ spectrum 
owing to the fact that it appears as a square of the amplitude.
The forward backward asymmetry, being proportional to the amplitude, is
reasonably large. It is about
$10\%$ in the $K^{*}$ resonance region where the number of events is maximum
as it can 
be seen in Fig.(5).

\begin{figure}
{\psfig{figure=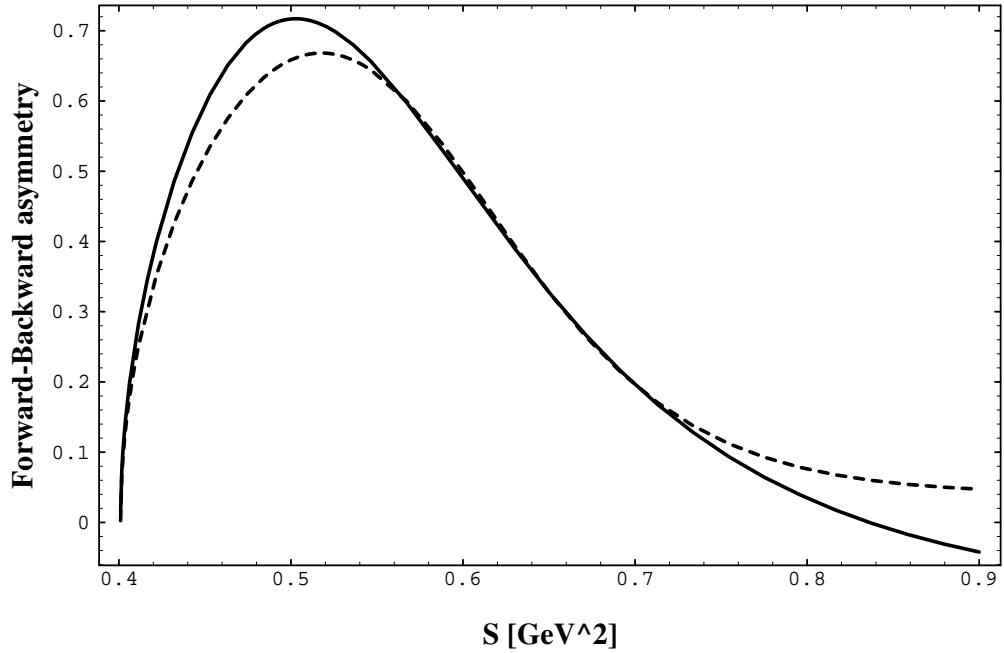,height=8.5truecm,angle=0}}
\caption{\em  Prediction for the forward-backward asymmetry $A_{FB}$ defined in eq(5) as 
a function of the $\pi K$ invariant mass squared. The dashed/solid curves
correspond respectively to the calculations
with/without the left hand cut of $\pi K$ scattering amplitude.}
\end{figure}

 The 
forward-backward asymmetry could be a useful quantity
for studying the relative phases of the S and P waves.

The Omn\`es representation with the S and P wave $I=1/2$ $\pi K$ scattering 
amplitude studied in Ref.\cite{belp} (and given in Fig.(6)) yields a branching ratio of
 $1.15\%$ for 
$\tau \to\pi K\nu$ decay, in a better agreement with 
the experimental value $1.4\pm 0.2 \%$.

\begin{figure}
{\psfig{figure=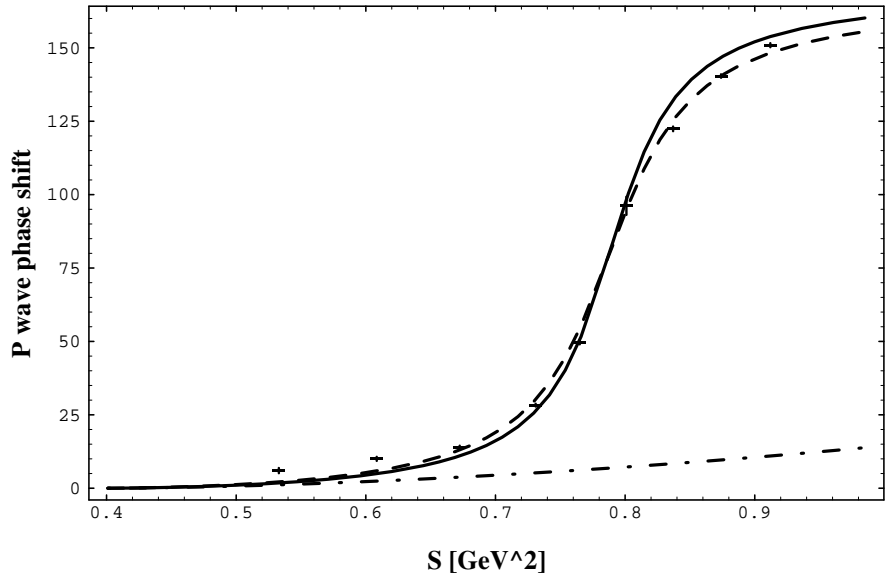,height=7.5truecm,angle=0}}
\caption{\em The solid line represents the I=1/2, l=1 $\pi K$ scattering phase shift
calculated from
the unitarized CPTh.
The dashed line corresponds to the similar phase shift when the left
 hand cut is neglected.
 The dot-dashed line is the CPTh prediction phase
(which is not the same
as the phase shift due to the violation of the full elastic unitarity relation 
in this
method).}
\end{figure}

\section{$\tau\to K\eta\nu$}
In this process elastic unitarity condition does not apply.
 One should consider the two channels $\pi K$ and $K\eta$ in the final 
state interaction. Chiral Perturbation Theory cannot describe the $K\eta$
 scattering even
at threshold since the $K^*$ resonance occurs below $K\eta$ threshold. 
According to the equivalence between the Pad\'e Method and the VMD approach 
(as it was pointed out in the pion form factor calculation), we will use in
the following the $K^*$ dominance for studying the $\tau\to K\eta\nu$ decay.
 VMD method
 introduces the inelastic effects through the $K^*$ self energy.
Using the exprimental value of $\tau\to K^*\nu$ branching ratio
 and SU(3) symmetry, we predict a branching ratio of $1.6\times 10^{-4}$. 
The experimental value of $\tau\to K\eta\nu$
 branching ratio is
 $(2.5\pm 0.5)\times 10^{-4} $.

\section{$\tau\to\pi\rho\nu$, $\tau\to\pi K^*\nu$ and $\tau\to K \rho\nu$}

 The most general
matrix element can be written as:
\begin{equation}
\langle\pi^{-}(k) \rho^{0}(p)\vert A^{1-i2}_{\mu} (0)\vert0
\rangle = f_1(Q^2)\epsilon_{\mu}+\epsilon .k\left( (k+p)_{\mu} f_2(Q^2)+
(k-p)_{\mu} f_3(Q^2) \right)
\end{equation}
where $Q^2=(k+p)^2$ and $\epsilon $ is the polarisation vector
 of $\rho$. $f_1$,
$f_2$, and $f_3$ are complex form factors and are only functions of $Q^2$.
Current
algebra soft pion theorem, which is obtained by taking the limit $k_{\mu}
\to 0$,
gives only information on $f_1$ but not on the other two form factors. In an
explicit model, it was shown that they contribute little to the $\tau \to
\pi \rho \nu $. Interested readers are refered to the original article
\cite{prehist}. (We assume here
that the decay constant of $\pi '$ is sufficiently small and hence can be
neglected). Using
the standard low energy current algebra theorem and taking the limit $ k_{\mu}
\to 0$ we have:
\begin{equation}
 \lim_{k_{\mu}\rightarrow 0} \langle\pi^{-}(k) \rho^{0}(p)\vert
A^{1-i2}_{\mu} (0)\vert0\rangle =-{\sqrt 2}{f_{\rho}\over f_{\pi}}\epsilon
_{\mu}(p)
\end{equation}
where $f_{\pi}=93MeV$, and $f_{\rho}$ is defined by the rate of $\rho \to
 e^{+} e^{-}$. Using the experimental data we obtain,
$f_{\rho}=0.118 GeV^2$. This
value of $f_{\rho}$ is equivalent to writing approximately the pion form
factor as\hfill\break
$F_{\pi}(s)= m_{\rho}^2(1+\delta s/m_{\rho}^2) / \left ( m_{\rho}^2-s-i
m_{\rho}\Gamma_{\rho}(s)\right)  $. A good fit to the
experimental data is obtained with $\delta =0.2$. In fact, the more general
form of
Eq(3) reads
 \begin{equation}
\lim_{k_{\mu}\rightarrow 0} \langle\pi^{-}(k) \pi^{+}(q_1)
\pi^{-}(q_2)\vert A^{1-i2}_{\mu} (0)\vert0\rangle =-{{\sqrt 2}\over
f_{\pi}}F_{\pi}(s)(q_1-q_2) _{\mu}
\end{equation}
 For convenience we shall first use
Eq(3). The $3\pi$  matrix element below the $\rho \pi$ threshold can be
straightforwardly obtained from Eq(28). 
 Using Eq(26) in (27) we have:
\begin{equation}
f_1(m_{\rho}^2)=-{\sqrt 2}{f_{\rho}\over f_{\pi}}
\end{equation}
Let us start with the narrow width approximation for the $A_1$ propagator.
Using $A_1$ dominance for
the form factor we have:
   
\begin{equation}
 f_{1}(Q^2) = -{\sqrt 2}{ f_{\rho}\over f_{\pi} } { (m_{A}^2-m_{\rho}^2) 
\over m_{A}^2-Q^2 }
\end{equation}
The generalisation of Eq[30] to take into account of the unstable nature of
$A_1$ can be
straightforwardly made. Using the $A_1$ dominance hypothesis for the axial
current, the general expression for $f_1(Q^2)$ is:
\begin{equation}
f_{1}(Q^2) = -{\sqrt 2}{ f_{\rho}\over f_{\pi} } { m_{A}^2-m_{\rho}^2-\pi
(m_{\rho}^2)  \over m_{A}^2-Q^2-\pi (Q^2) }
\end{equation}
where we use the standard prescription for describing an unstable
 particle, with
$\pi (Q^2)$ being the self energy operator of the $A_1$ resonance and is
obtained by the
bubble summation of the $\pi \rho$ intermediate states, similar to the
treatment of the W and Z propagators in the standard model. In order to have
the usual Breit Wigner description of a resonance, we must make a twice
substracted dispersion relation with $Re[\pi(m_{A}^2)]= Re[\pi
'(m_{A}^2)]=0$ where
$m_A$ is the $A_1$ mass:

\begin{eqnarray}
&Re[\pi (Q^2)]\hskip -1.5mm=\hskip -5mm& {(Q^2-m_A^2)^2\over
\pi}P\hskip -8mm\int\limits_{(m_\pi+m_{\rho})^2}^{\infty}\hskip -8mm
dz{Im[\pi
(z)]\hskip -1.2mm -\hskip -1.2mm Im[\pi(m_A^2)]\hskip -1.2mm -\hskip -1.2mm (z-m_A^2)Im[
\pi ' (m_A^2)]\over (z-m_A^2)^2(z-Q^2)}\cr
&Im[\pi (Q^2)]&={ g_{A\rho\pi}^2\over 8\pi } {{\sqrt
{\lambda (Q^2,m_{\rho}^2,m_{\pi}^2)}}\over Q^2}\left(1+{\lambda
(Q^2,m_{\rho}^2,m_{\pi
}^2) \over 12m_{\rho}^2 Q^2}\right)
\end{eqnarray}

where we define the $\pi^{0} \rho ^{+} A_1^{-}$ vertex as
$g_{A\rho\pi}\epsilon (A)
.\epsilon (\rho)$, $\lambda
\left(Q^2,m_{\rho}^2,m_{\pi}^2\right)=(Q^2-(m_{\rho}+m_{
\pi})^2)(Q^2-(m_{\rho}-m_{\pi})^2)$, and P stands for the principal part
integration.

If we assume that the acceptance
correction to the ALEPH data was negligible, our best fit to the $3\pi$
spectrum gives the following
ranges of $m_{A}$ and $\Gamma _{A}$: $m_A=1.24\pm 0.02 GeV$, $\Gamma
_A=0.43\pm 0.02$ GeV.

Using these results, the calculated branching ratio for $\tau\to 3\pi \nu$
is $19 \pm 3 \%$. The
central value corresponds to our best fit which is shown in Fig.(7). This
value agrees with 
the ALEPH branching ratio $19.14\pm 0.48\pm0.44\%$.

\begin{figure}
{\psfig{figure=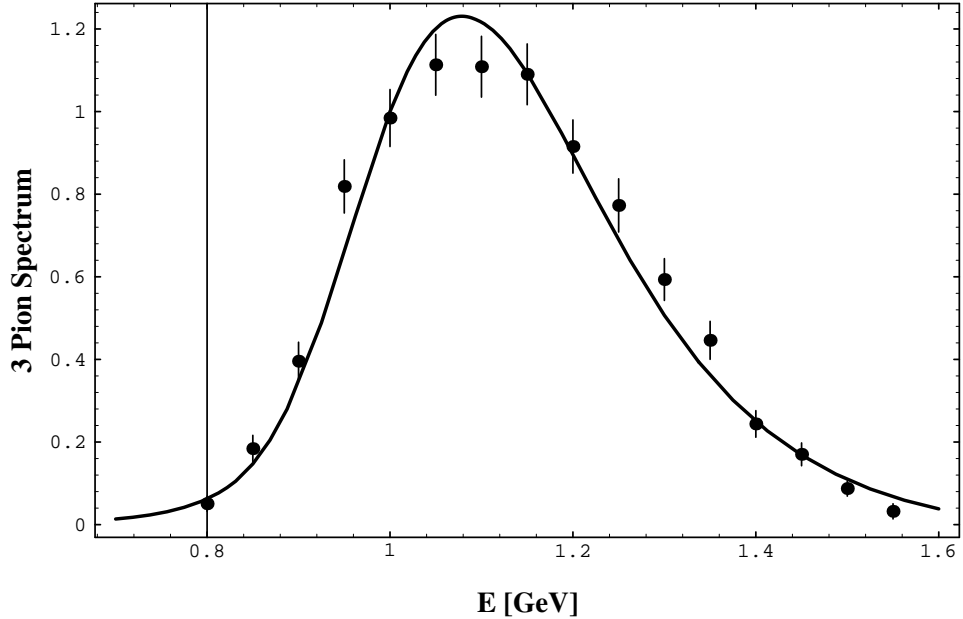,height=8truecm,angle=0}}
\caption{\em  Our best fit for the $\tau \to 3\pi \nu$
spectrum corresponding to $m_A=1.24$ GeV, $\Gamma_A=0.43$ GeV}
\end{figure}

A similar treatment for $\tau\to\pi K^*\nu$ and $\tau\to K\rho\nu$ decays
 can be done using the  axial vector dominance of
 $Q_1(1270)$ and $Q_2(1400)$ (for more details see Ref.\cite{bels}).
Numerical calculation gives:
\begin{eqnarray}
BR(\tau \to \rho^{0} K^{-}\nu)=0.1\% \nonumber\\
BR(\tau \to \pi^{-} K^{0*}\nu)=0.4\%
\end{eqnarray}
These results are in good agreement with the values of TPC$/$Two-Gamma
collaboration: $B\left (\tau \to K^{* 0}\pi^{-}\nu,neutrals\right)  =
0.51\pm 0.2\pm 0.13
$ and \\$B\left (\tau \to K^{-}\pi^{+}\pi^{-}\nu\right) =0.7 \pm 0.2$. The
$K^{-}\rho^{0}\nu$ mode is therefore consistent with zero.

\section{Conclusions}

To conclude we would like to outline some main points:

{\it (i)} Perturbation theory satisfies unitarity order by
order and that is known as perturbative unitarity.

This method is sufficient to describe physical phenomena if the
 coupling constant is small like in QED.
But it fails if there is a bound state or a resonance state: in QED the positronium bound state is solved by Bethe-Salpeter equations (Ladder type summation).

{\it (ii)}-- Notice that in the Standard Model, the unitarity is preserved by the bubble summation of the $Z^0$ propagator.

--In the Nambu-Jona-Lasinio Model the bubble summation is done
 to satisfy the unitarity. The Higgs and Nambu Goldstones are generated as bound states due to four fermion interaction.

{\it (iii)} Low energy Nucleon-Nucleon interaction has resonance in the
singlet S state and a bound state
 (Deuteron) in the triplet state. This problem was solved by
H. Bethe in the 50's using effective range expansion which preserves the
unitarity of the S matrix. More explicitly, instead of expanding
 the amplitude like in Chiral Perturbation Theory, the   Effective Range Theory
makes an expansion of $k\cot (\delta)$
\begin{equation}
k\cot (\delta)= {1\over a} +{1\over 2}r_0 k^2+...
\end{equation}

with $a$ and  $r_0$ being respectively the scattering length
 and the effective range.

This expansion, unlike $\chi$PT, preserves unitarity.

{\it (iiii)} In this paper we have shown that Chiral Perturbation
 Theory should be modified to describe the main problems of $\tau$ physics.
We have extended the range of validity of Chiral Perturbation Theory 
using the Pad\'e Approximant Method or VMD hypothesis. Both methods satisfy
unitarity. A good agreement with the experimental data is obtained 
for the exclusive hadronic modes of $\tau$ decay.

\end{document}